\documentclass[%
 reprint,
 a4paper, 
 amsmath,amssymb,
 aps,
 floatfix,
 superscriptaddress,longbibliography 
]{revtex4-2}
\usepackage{graphicx}
\usepackage{dcolumn}
\usepackage{bm}

\usepackage{xcolor}
\usepackage[toc,page]{appendix}
\usepackage{xcolor}
\usepackage{todonotes}
\usepackage{comment}

\usepackage{todonotes}

\begin{document}

\title{Lorentz-invariant topological structures of the electromagnetic field\\
in a Fabry-Perot resonant slit-grating}

\author{Marina Yakovleva}
\email{marina.yakovleva@c2n.upsaclay.fr}
\affiliation{%
Universit\'e Paris-Saclay, CNRS, Centre de Nanosciences et de Nanotechnologies, 91120, Palaiseau, France.
}%

\author{Jean-Luc Pelouard}%
\affiliation{%
Universit\'e Paris-Saclay, CNRS, Centre de Nanosciences et de Nanotechnologies, 91120, Palaiseau, France.
}%
\author{Fabrice Pardo}%
\email{fabrice.pardo@c2n.upsaclay.fr}
\affiliation{%
Universit\'e Paris-Saclay, CNRS, Centre de Nanosciences et de Nanotechnologies, 91120, Palaiseau, France.
}%

\date{\today}

\begin{abstract} 

It is commonly assumed that
the most correct description of the electromagnetic world is the abstract one,
and that topological constructs
such as lines of force
are not
covariant.
In the present paper, we show that
for a $y$-invariant system
with a $p$-polarized electromagnetic field,
it is possible to construct
absolute (i.e. Lorentz invariant) 
lines, which we call \emph{electric spaghettis} (ESs).
The electromagnetic field is fully described
by the ES topology that transcend the limit between space and time,
plus a new invariant,
the characteristic parameter $\eta$.
In a Fabry-Perot resonant slit-grating,
three ES patterns can be distinguished,
corresponding to three regions:
null straight lines in plane wave regions,
rounded rhombuses 
in the interference region
inside the grating,
and Bernoulli's logarithmic spiral patterns --- the first ever described fractals --- in the funneling region.
\end{abstract}



\maketitle

Faraday's concept of lines of force \cite{faraday1852physical,maxwell1861opl} was
the first topology describing
the electromagnetic field.
However, with our modern point of view,
this topological construct is not Lorentz invariant, and it is commonly assumed that
\textit{the most correct description of the electromagnetic field is the abstract one} \cite{feynman1977feynman, dyson2007wim}.
Nevertheless, few attempts to define a spacetime topology of the electromagnetic field have been made.
The idea of tubes of flux, a spacetime generalization of Maxwell's concept \cite{maxwell1861opl}, was described for example by
Misner, Thorne and Wheeler in their book \textit{Gravitation} \cite{misner1973gravitation}.
Unfortunately it only leads to 
a schematic presentation and does not construct well-defined spacetime topological structures.
A frequent approach of field 2-forms
is to separate space and time \cite{deschamps1981electromagnetics,warnick2014differential}, losing the Lorentz invariance.
Absolute representations of 2-forms were
given only for simple cases,
by Jancewicz for an electric charge \cite{bernard1989multivectors} and by Warnick and Russer for the case of a propagating plane wave \cite{warnick2006two}.

In an attempt to construct a Lorentz-invariant topology of the electromagnetic field,
Gratus \cite{gratus2017pictorial} has suggested to consider fixed-coordinate space slices
of the 4D spacetime.
A space slice, being arbitrary in the general case,
is canonical for a $y$-invariant $p$-polarized electromagnetic field.
In this case, the Maxwell's 2-form becomes a simple 2-form, and its pullback onto any arbitrary $y = \mathrm{Constant}$ plane defines lines in the 3D spacetime.
There is similarity between these lines and
Faraday's lines of force for a static field in
3D space. The topology by itself defines at any point (or event) two directional parameters,
and a third parameter
is necessary to completely describe the field,
either the value of the electrostatic potential,
or the field amplitude (that can be illustrated
by the density of lines).
However, this field description by a topology plus a parameter
cannot be straightforwardly generalized from 3D space to 3D spacetime, mainly because of the peculiarity of null-like directions in the latter.
In the absence of charge, it is evident that field lines never end in 3D space, but this is not
an established fact in 3D spacetime.

In this paper, we show that a $y$-invariant,
$p$-polarized electromagnetic field
can be fully described by an absolute topology
of lines in spacetime (named here the \emph{electric spaghettis} (ESs)),
with a new Lorentz-invariant scalar measure on them, the $\eta$ parameter.
This topology is studied in detail for
an infinite line of charge,
and for a Fabry-Perot resonant slit grating.
The latter exhibits three specific topological ES structures: the single plane wave region high above the grating, the two interfering plane wave regions inside the grating, and the funnel region in the near field of the grating.
In the funneling region a fractal topological structure
around null field events is observed.
These endless ES whirls,
which seem to contradict the absence of magnetic charge \cite{gratus2017pictorial},
are justified here by a careful analysis.

For $y$-invariant $p$-polarized field, the 6 components of Faraday's $F = (\bm{-E},\bm{B})$ or Maxwell's $\mathcal{G} = (\bm{H},\bm{D})$ 2-forms \cite{stern2015geometric} reduce to 3 components, with
\begin{equation}
\mathcal{G} = H_y dt \wedge dy +  D_x dy \wedge dz + D_z dx \wedge dy.
\label{eq:G}
\end{equation}

This 2-form can be written as
$\mathcal{G}
    = \phi_e \wedge dy$, with $\phi_e$ the 1-form
\begin{equation}
    \bm{\phi}_e = H_y dt  + D_z dx - D_x  dz.
    \label{eq:ElFluxes}
\end{equation}
The dual of this 1-form is the vector: 
\begin{equation}
    \bm{\tau}_{S} =  H_y \partial_t  - D_z \partial_x +  D_x \partial_z.
    \label{eq:tau}
\end{equation}
This vector defines integral curves, the ESs,
that are absolute topological structures in 3D spacetime:
a given line connects the same set of events, independently
of any frame of reference.
Starting from from any event $\bm{s}_0$,
the line $\mathrm{ES}(\bm{s}_0)$ can be built using the parametric equation
\begin{equation}
\bm{s}(\eta) = \bm{s}_0 + \int_{0}^{\eta} \bm{\tau}_S d\eta . 
\label{eq:dseta}
\end{equation}
The electric flux on a spaghetti segment $\mathrm{ES}_{e_1e_2}$ from the event
$\mathrm{e}_1 = \bm{s}(\eta_1)$ to the event
$\mathrm{e}_2 = \bm{s}(\eta_2)$ has the value
\begin{equation}
 \varphi_e (\mathrm{ES}_{e_1e_2}) = \int_{\eta_1}^{\eta_2}  (H^2 - D^2)  d \eta.
 \label{eq:invarFlEt}
\end{equation}
In this equation, the electric flux value $\varphi_{e}$ and the
$H^2 - D^2$ factor are both Lorentz-invariant. 
Hence $\eta(\mathrm{ES}_{e_1e_2}) = \eta_2 - \eta_1$
is actually a new Lorentz invariant.

From a topological point of view, a set of ESs can be
constructed to cover any region of the 3D spacetime.
Furthermore, an oriented segment of ES 
can be labeled with three quantities:
1) its interval length
(but only if the segment is purely timelike or spacelike),
2) its electric flux $\varphi_{e}$,
3) its characteristic parameter $\eta$.

Unlike worldlines of physical objects,
which are always timelike,
ESs transcend the boundary between time and space directions.
Some parts of the same ES can be timelike, spacelike, and even lightlike.
The particular case
where ESs are contained in a light cone
requires some attention, as
$H^2 - D^2$,
the interval and the electric flux
are all equal to zero.
In contrast,
the $\eta$ value of a non-degenerate (i.e. not reduced to a point) ES segment is never zero, and has a sign fixed by the orientation of ES.
If an arbitrary inertial frame of reference is chosen,
the components of the electromagnetic field can be obtained
from the ES topology and the $\eta$ value using Eq.~(\ref{eq:dseta}):
\begin{equation}
H_y = \frac{ds_t}{d\eta},  D_z = -\frac{{ds_x}}{d\eta},  D_x  = \frac{{ds_z}}{d\eta},
\label{eq:dsdeta}
\end{equation}
where $s_t, s_x, s_z$ are the ES coordinates.
The smaller the field, the larger the $\eta$ value.
It may seem rather counter-intuitive,
but the ES characteristic parameter
$\eta$ is an absolute, extensive physical property of ESs.

\begin{figure}
	\center{
	  \includegraphics[width=\linewidth]{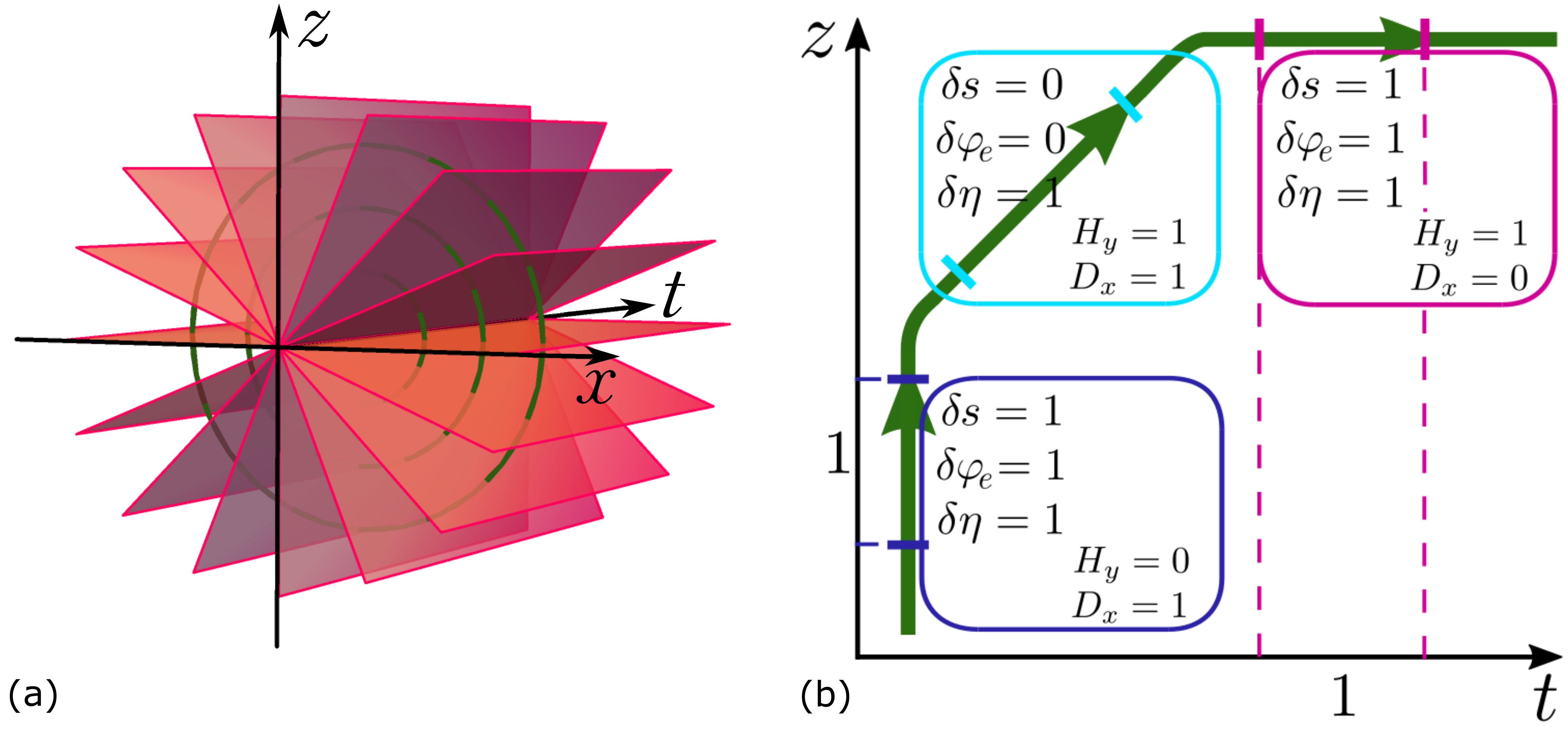}
	}
	\caption{(a) A $t$-invariant (electrostatic) case: an $Oy$ charged line with $Ot$ worldline. Green circles: ESs, pink surfaces: ZEFSs.
	(b) A non-$t$-invariant case: ES (green line) with purely electric field part ($z$ direction, dark-blue frame), plane-wave part (null-like direction, light-blue frame), and purely magnetic field ($t$ direction, purple frame). 
	The ES topology is absolute, the interval values $\delta s$,
	the flux values $\delta \varphi_e$ and the characteristic parameter values $\delta \eta$ are all absolute, independent of the reference frame. In contrast, $H_y$ and $D_x$ depend on the axis choice ($t, z)$, with Eq.~\ref{eq:tau} and Eq.~\ref{eq:dsdeta}. When ES is null-like, only the  parameter $\eta$ has a non-zero value,
	able to describe the electromagnetic field in this region.}
	\label{fig:chmap}
\end{figure}

Fig.~\ref{fig:chmap}(a) shows the case of a linear, constant density of charge $\lambda$ along the $y$-direction.
The worldline of each charge is a straight line
along the $Ot$ axis,
and this axis represents all charges in 3D spacetime
$(t, x, z)$.
ESs appear as concentric green circles centered around the $Ot$ line.
Every ES circle contains the same electric flux value
$\varphi_e = \lambda$.
The pink radial surfaces are Zero-Electric Flux Surfaces (ZEFSs).
They are perpendicular to the ESs and start on the $Ot$ charge worldline. 
The electric flux on any line they contain is zero.

At first glance,
Fig.~\ref{fig:chmap}(a) is a picture of equipotential lines (green circles, which are indeed cylinders along $y$-axis in 4D spacetime) 
and electric field strength lines, extended in the time direction (pink surfaces).
However, the concept of equipotential lines is frame-dependent:
in a different frame of reference, 
a magnetic field would appear,
and both the electric field
and the equipotential lines would change.
In contrast, ESs and ZEFSs are absolute,
connecting the same set of events,
independently of any frame of reference.
The $\eta$ value calculated between two events of a given ES is absolute,
independent of the choice of a reference frame.
The field components, which are relative, can be simply derived from the ES topology and the parameter $\eta$,
using Eq.~\ref{eq:dsdeta}.

In this electrostatic example, $\varphi_e$  has the same value
$\lambda$ on all circles.
By contrast, $\eta = 4 \pi^2 R^2/\lambda$ is proportional
to the square of the radius $R$.
The usefulness of the parameter $\eta$,
not obvious in this example, is crucial
when the ES part belongs to a light cone,
as shown in Fig.~\ref{fig:chmap}(b),
where an ES is successively
space-like, null-like, and time-like,
transcending the distinction between space and time.

\begin{figure*}[tb]
	\center{
	    \includegraphics[width=0.9\linewidth]{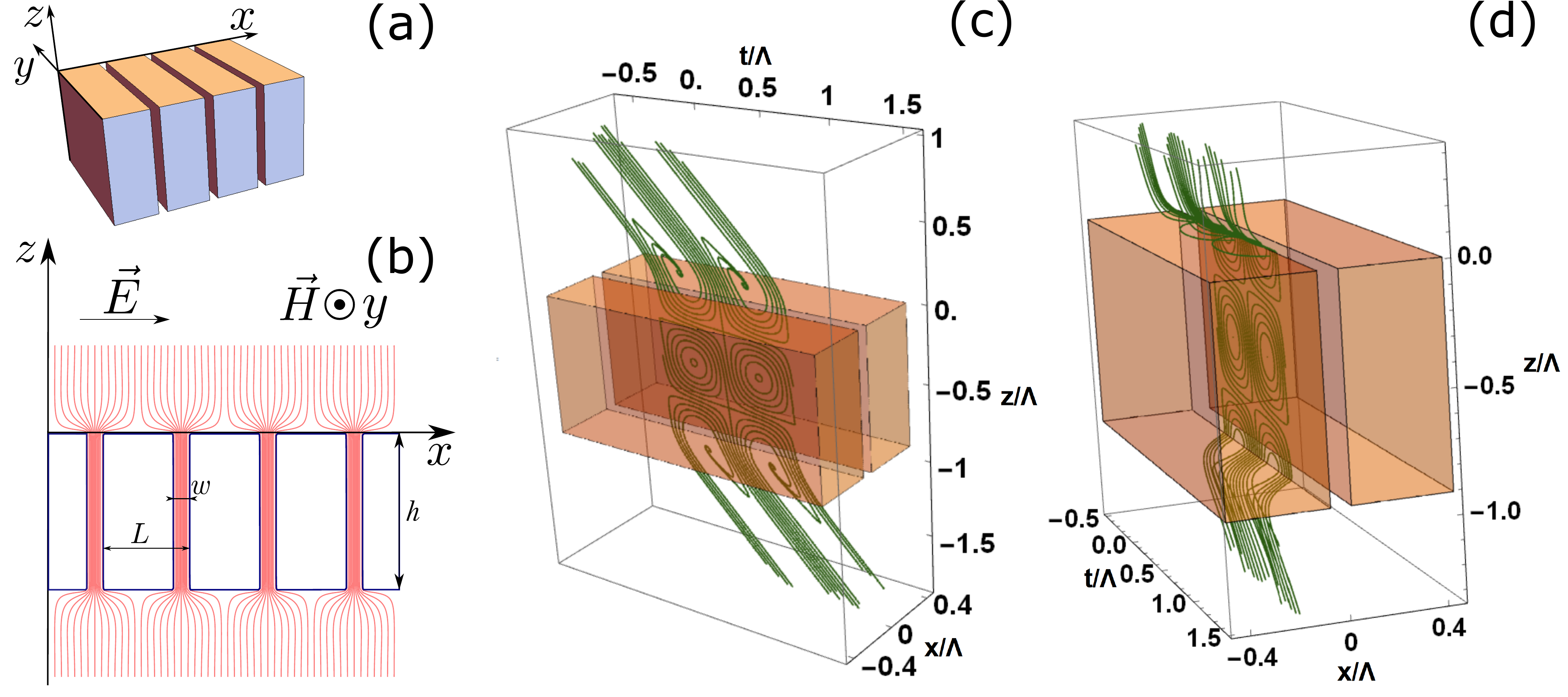}
	}
	\caption{(a) Slit grating,
	invariable in $y$-direction, made in a perfect conductor.
	(b) Time-averaged Poynting vector lines
	in the resonsant case, $p$-polarized normally incident monochromatic wave.
	(c) ESs in the central plane of a slit.
	(d) ESs crossing the line \textit{z = - h/2, x =- 2 w/5}.}
	\label{fig:gratSpag}
\end{figure*}

The Fabry-Perot resonant slit grating system studied in this paper
consists of a perfectly conducting metal with slits parallel to the $y$-direction (Fig.~\ref{fig:gratSpag}(a)),
excited at normal incidence
by a \textit{p}-polarized plane wave
(i.e. with the magnetic field oriented along the $y$ direction).
The grating period is $L = \pi$,
that is, half the wavelength $\Lambda = 2 \pi$ ($c = 1$),
so all the diffracted waves are evanescent.
The slit width is arbitrarily chosen as $w = L/6$.
This system is known for perfect transmission of incident radiation \cite{stavrinou2002propagation},
with a funneling mechanism (Fig.~\ref{fig:gratSpag}(b)) that
can be understood by the magneto-electric interference
between the incident and evanescent fields \cite{pardo2011light}.
The Fabry-Perot resonance condition corresponding to
total transmission and no reflection
is obtained for a slit height $h$ equal
to $h = h_0 + n \, \Lambda /2$
with $h_0 \approx 0.417 \, \Lambda$,
and $n$ is an arbitrary integer.
The value of $h_0$ is slightly smaller than $\Lambda/2$,
as the phase
of the internal reflection
is not exactly $\pi$.
The value $h = 0.917 \, \Lambda$
was chosen for a clear view of the interference
pattern inside the slits.

\begin{figure}
	\center{
	    \includegraphics[width=0.9\linewidth]{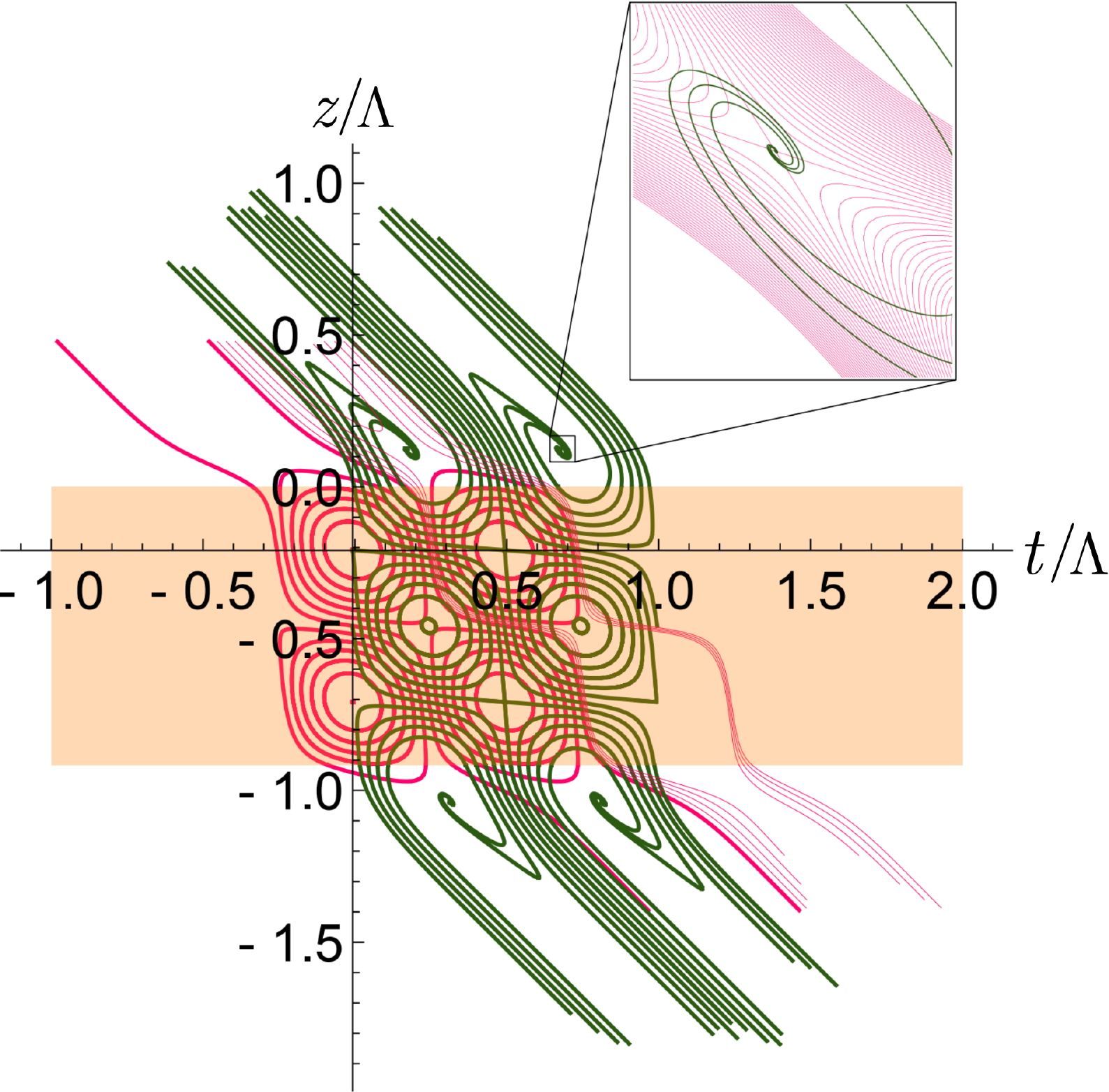}
	}
	\caption{Plot of the ES (green)
	and ZEFS (red) lines on the median plane
	for one time period.
	The orange zone corresponds to the slit area.
	The electric flux is constant between two thick
	red ZEFS lines, and also between two thin red lines,
	with a factor $1/6$.
	Inset: zoom of the whirl zone, with very small
	flux step between ZEFS lines.}
	\label{fig:spCent}
\end{figure}

Several numerical techniques based on Eq.~\ref{eq:dseta} and Eq.~\ref{eq:invarFlEt} were used to compute
ESs and ZEFSs.
Full descriptions of these and codebooks
are presented in Supplemental Material.
Fig.~\ref{fig:gratSpag}(d) shows ESs
crossing the line $z = - h/2, x=-2w/5$.
The median plane of a slit is a special place
because the $D_z$ component is here equal to zero,
and the ESs remain localized in this plane (see Eq.~\ref{eq:tau} and Fig.~\ref{fig:gratSpag}(c)).
These ESs are also plotted in Fig.~\ref{fig:spCent},
along with ZEFS lines.
As there are no electric charge in the considered region,
the electric flux is constant between
two
neighboring bold ZEFS lines,
and
the density of these lines 
in the $t$-direction
corresponds to the value of the field component $H_y$,
and the density in the $z$-direction
corresponds to the value of the field component $D_x$.

Far from the slits, where a single plane wave propagates,
$D_x = H_y$, the ESs are straight null-like lines,
according to Eq.~\ref{eq:tau}.
Inside a slit, where two plane waves interfere,
concentric patterns of rounded rhombi are observed.
The field is purely electric when the ESs (green lines)
are parallel to the $z$-direction,
and purely magnetic when they are parallel to the $t$-direction.
Owing to the resonance inside the slit,
the electric flux density in this region
is six times higher
than in the plane-wave region.
This is clearly depicted in Fig.~\ref{fig:spCent}:
in the slit region, there are six ZEFS lines
over a half-period in time,
whereas in the upper part of the figure
(plane-wave region), there is only one bold ZEFS line.
Compared with the interference of two identical plane waves \cite{pardo2016seminar},
the rounded squares
are slightly distorted.
This is because
the amplitude of the wave from bottom to top 
is smaller than that from top to bottom.

\begin{figure}
	\center{
	    \includegraphics[width=\linewidth]{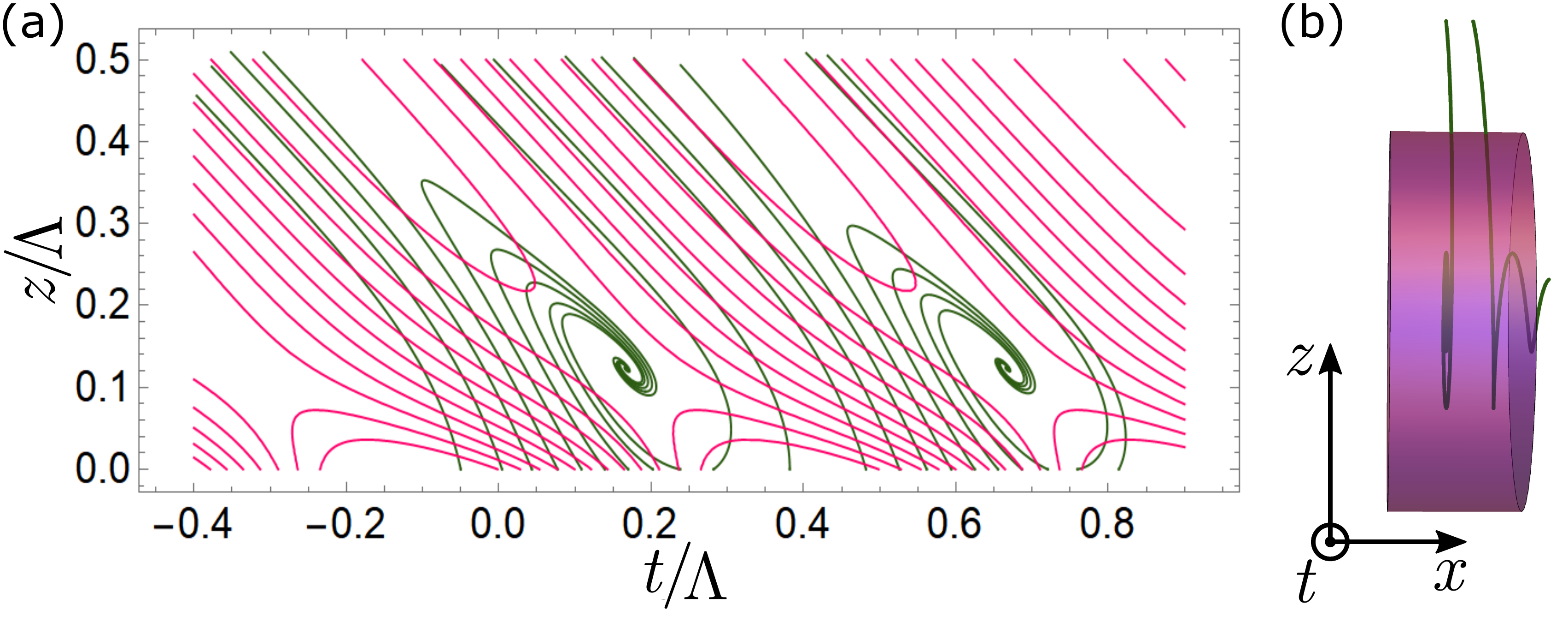}}
	\caption{(a) ESs (green) and ZEFS-lines (red) in the case of interference of the plane wave and two evanescent ones (with equal amplitudes and opposite propagation directions along $x$ axis).
	(b) Spacetime camembert-box with incoming, never outgoing ESs at $x = 0$. Other ESs are outgoing through one of the sides.}
	\label{fig:intoftwo}
\end{figure}

In the funneling region, ESs form whirls 
that are centered on zero field events.
They seem to contradict the principle
that tubes of flux with no charge never end \cite{misner1973gravitation},
so they require detailed analysis.
The field in the funneling region is the superposition
of the incoming plane wave and of a series of diffracted evanescent waves.
For simplicity, 
let us
restrict the continuation of the study to only the first diffracted wave.
In this model, the incident plane wave has amplitude $\cos(z + t)$,
and the evanescent wave has amplitude
$a \exp{( - \sqrt{3} z)} \sin(t) \cos(2 x)$
where the amplitude $a = 1.091$ and the phase
$\pi/2$ are obtained numerically
(see Supplemental Material).
The ES-ZEFS pattern of this interference
is plotted as respectively green and red lines in Fig.~\ref{fig:intoftwo}(a).
The pattern of this simplified model is similar to 
the one observed in Fig.~\ref{fig:spCent}.
The whirls are centered on zero-field points $H_y = D_x = 0$.

Using a linear approximation of the field components
around the zero-field points $(t_0 = \pi /3 + p/2, \; z_0 = 0.7736)$,
the parametric equation of an ES is (see Supplemental Material):
\begin{align}
  \begin{pmatrix}
    t(\eta)-t_0\\z(\eta)-z_0
  \end{pmatrix}
  &=
  z_0
  \begin{pmatrix}
    0 & \beta \\
    1 & -\alpha
  \end{pmatrix}
  \begin{pmatrix}
    \cos {\lambda_i \eta} \\
    \sin {\lambda_i \eta}   
  \end{pmatrix}
  e^{-\lambda_r \eta}
  \label{eq:defspirtz}\\
  x(\eta) &= x_0 e^{2 \lambda_r \eta} \label{eq:defspirx}
\end{align}
where $ \lambda_i = \sqrt{(3 \sin{z_0} + \frac{\cos{z_0}}{\sqrt{3}})(\sin{z_0} - \frac{\cos{z_0}}{\sqrt{3}})}$, and
$\lambda_r =  \sin{z_0} -  \frac{\cos{z_0}}{\sqrt{3}}$.
For $x_0 = 0$ this is an affine transformation of a curve, named \emph{spira mirabilis} by Bernoulli three centuries ago \cite{hammer2016spira}.
It is an endless spiral,
the first fractal curve ever studied.
And this is the key of the paradox that
ES should never end. They never end indeed.
Each loop of ES toward the center corresponds
to $2 \pi/\lambda_i$ change of the parameter $\eta$
while the radius decreases by a factor of $\exp(-2 \pi \lambda_r/\lambda_i) = 0.12$.

The ES bundles make tubes of magnetic flux,
corresponding to the Faraday 2-form $(-\bm{E},\bm{B})$,
with $E_x dx \wedge dt + E_z dz \wedge dt + B_y dx \wedge dz$. 
As there are no magnetic charges, any tube tube entering on a given closed surface $\Sigma$ of spacetime $(t, x, z)$ must exit somewhere on the same surface. 
The observed endless ESs do not contradict this law because they are restricted to $x = 0$ plane, having zero measure on the surface $\Sigma$. 
If we consider a camembert-box as in Fig.~\ref{fig:intoftwo}(b) the incoming ESs at the periphery exit through the sides of the disk thus defining tubes of constant magnetic flux.

To conclude, we demonstrated that
a $y$-invariant \emph{p}-polarized
electromagnetic field,
which is usually represented by 3 field components
(as $H_y, D_z, D_x$, relative to a frame of reference),
can be represented in an absolute way.
In a flat spacetime, the qualifier \emph{absolute}
is synonymous with \emph{Lorentz invariant}.
It should be noted that the content of this letter can be extended
to a curved spacetime, 
provided that there is translation symmetry along $y$ direction.
The electromagnetic field
can be represented by unique and absolute structures,
the electric spaghettis (ESs),
measured by the absolute parameter $\eta$.
The surfaces perpendicular to ESs (the Zero Electric Flux Surfaces (ZEFSs)) are also absolute space-time structures.
Contrary to the worldlines of physical objects,
which are always timelike,
ESs and ZEFSs transcend the limit between space and time.
Any segment of ES is characterized by
two absolute electromagnetic quantities:
its electric flux $\varphi_e$, and its characteristic parameter $\eta$.
Despite its counter-intuitive nature (the larger it is, the smaller the field),
$\eta$ is a parameter that makes sense because
it can describe the electromagnetic field everywhere.
In contrast, the flux and interval values,
which are zero on null-like segments of ESs, cannot describe the electromagnetic field everywhere.
ESs with $\eta$ milestones written on them,
form a complete representation
of the electromagnetic world in the 3D $y$-invariant spacetime, giving
in any inertial frame of reference $(t, x, z)$, 
the field components $H_y, D_z, D_x$.

We explored the ES topology
of a Fabry-Perot resonant slit-grating.
In the plane wave region ESs are straight light-lines, with an electric field equal to the magnetic field.
Inside the slits, the ESs are mainly rounded rhombi,
corresponding to the interference
of two plane waves, with purely $H_y$
zones and purely $D_x$ zones.
The funneling region is of particular interest,
with ESs having the fractal behavior of
logarithmic spirals.
They correspond to
the interference of the incident wave with
the evanescent waves diffracted by the slit-grating.

ESs illustrate
the profound unity of electric and magnetic fields,
and give them a topological structure
that can be studied for itself, opening a new
field of research.
Also, instead of considering field components,
which are related to an arbitrary frame of reference, 
the description of the field as an absolute
topology in spacetime with absolute measure $\eta$
is a radical change of the conceptual approach
to the electromagnetic world.
This not only has philosophical and educational consequences, but also suggests new computational approaches.

\bibliography{main}
\end{document}


\title{Supplementary materials to the letter\\
Lorentz-invariant topological structures of the electromagnetic field,\\
the Fabry-Perot resonant slit-grating case.}

\author{Marina Yakovleva}
\email{marina.yakovleva@c2n.upsaclay.fr}
\affiliation{%
Universit\'e Paris-Saclay, CNRS, Centre de Nanosciences et de Nanotechnologies, 91120, Palaiseau, France.
}%

\author{Jean-Luc Pelouard}%
\affiliation{%
Universit\'e Paris-Saclay, CNRS, Centre de Nanosciences et de Nanotechnologies, 91120, Palaiseau, France.
}%
\author{Fabrice Pardo}%
\email{fabrice.pardo@c2n.upsaclay.fr}
\affiliation{%
Universit\'e Paris-Saclay, CNRS, Centre de Nanosciences et de Nanotechnologies, 91120, Palaiseau, France.
}%

\date{\today}

\maketitle

\section{\label{sec:whirl}Calculation of whirl}

The flux and field components of the interference of a plane wave and the first stationary evanescent wave diffracted by a grating with period $L = \Lambda/2$ are
\begin{align}
  \varphi_e &= \sin{(z + t)} - a  \exp{(-\sqrt{3} z)} \cos{t} \cos{2 x} \label{eq:phiwhirl} \\
  H_y &= \cos{(z + t)} +  a \exp{(-\sqrt{3} z)}  \sin{t} \cos{2x} \label{eq:hywhirl} \\
  D_x &= -\cos{(z + t)} - a \sqrt{3}  \exp{(- \sqrt{3} z)} \cos{t} \cos{2x} \label{eq:dxwhirl} \\
  D_z &= 2 a \exp{(-\sqrt{3} z)} \cos{t} \sin{2 x} \label{eq:dzwhirl}
\end{align}
with $a = 1.091$ for the resonant slit grating we consider.

\begin{figure}
	\center{\includegraphics[width=0.9\linewidth]{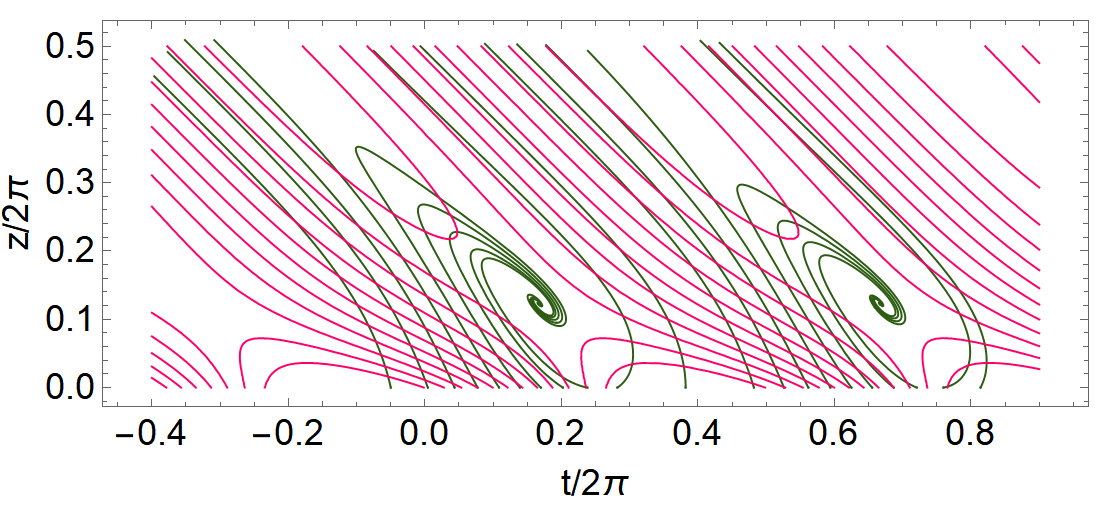} }
	\caption{ESs (green) and ZEFS-lines (red) in the case of interference of the plane wave and two evanescent ones(with equal amplitudes and opposite propagation directions along $x$ axis.).}
	\label{fig:intoftwo}
\end{figure}

The ES-ZEFS pattern of this interference
at $x = 0$ (corresponds to the middle plane of a slit for the grating described above)
is plotted as green and red lines in Fig.~\ref{fig:intoftwo}.
The pattern of this simplified model
is very similar to the one observed in Fig.~4 of the main text.

The whirls in Fig.~\ref{fig:intoftwo} are centered on zero-field points $H_y = E_x = 0$,
corresponding to equations:
\begin{align}
    \tan{t_0} &= \sqrt{3} \\
    \frac{\cos{z_0}}{\sqrt{3}} - \sin{z_0} + a \exp(-\sqrt{3} z_0) &= 0
\end{align}
with solution $(t_0 = \pi /3,  z_0 = 0.7736)$.

The asymptotic character of the whirls can be obtained by a linear approximation
$(t = t_0 + t',  x = x', z = z_0 + z')$ of the field around the point $(t_0, 0, z_0)$:
\begin{align}
  \begin{pmatrix}
    H_y\\D_x
  \end{pmatrix}
  &= \lambda_i A
    \begin{pmatrix}
    t'\\z'
  \end{pmatrix} \label{eq:linhydx} \\
  D_z &= \lambda_x x' \label{eq:lindz}
\end{align}
with
\begin{align}
  A &=  \begin{pmatrix}
    -2 \alpha + \beta & -\beta \\
    \beta & \beta
  \end{pmatrix} \\
  \lambda_i &= \sqrt{(3 \sin{z_0} + \frac{\cos{z_0}}{\sqrt{3}})(\sin{z_0} - \frac{\cos{z_0}}{\sqrt{3}})} \\
  \lambda_i \alpha &= \sin{z_0} +  \frac{\cos{z_0}}{\sqrt{3}} \\
  \lambda_i \beta &= 2 \sin{z_0} \\
  \lambda_x &= \lambda_i (-2 \alpha + 2 \beta).
\end{align}
The factor $\lambda_i$ will appear later as the angular frequency for ES $\eta$ parameter.
From these equations we obtain
\begin{equation}
  \beta^2 = 1 + \alpha^2.
\end{equation}

Applying Eq.(6) from the main part, ES equations becomes:
\begin{align}
  \frac{d}{d\eta}
  \begin{pmatrix}
    t'\\z'
  \end{pmatrix}
  &= \lambda_i A
  \begin{pmatrix}
    t'\\z'
  \end{pmatrix} \label{eq:diffA} \\
  \frac{d x'}{d\eta} &= -\lambda_x x.
\end{align}
The solution of this system of differential equations is
\begin{align}
  \begin{pmatrix}
    t'(\eta)\\z'(\eta)
  \end{pmatrix}
  &= e^{\lambda_i \eta A}
  \begin{pmatrix}
    t'(0)\\z'(0)
  \end{pmatrix} \\
  x'(\eta) &= e^{- \lambda_x \eta} x'(0)
\end{align}
where the exponential of matrix $\lambda_i \eta A$
can be obtained from its eigenvalues $\lambda_{\pm}$
and eingenvectors matrix $V$:
\begin{equation}
    \exp(\lambda_i \eta A) = 
    V
    \begin{pmatrix}
    \exp(\lambda_+ \eta) & 0 \\
    0 & \exp(\lambda_- \eta)
    \end{pmatrix}
    V^{-1}
    \label{eq:expAeta1}
\end{equation}
where
\begin{equation}
    \lambda_i A V  = 
    V \begin{pmatrix}
    \lambda_+ & 0 \\
    0 & \lambda_-
    \end{pmatrix}.
\end{equation}
Explicitly
\begin{equation}
    \lambda_\pm = \lambda_r \pm i \lambda_i
\end{equation}
with
\begin{equation}
    \lambda_r = (\beta - \alpha) \lambda_i = \lambda_x/2\\
\end{equation}
and
\begin{align}
  V &= \begin{pmatrix}
    -\alpha + i & -\alpha - i \\
    \beta & \beta
  \end{pmatrix} \\
  V^{-1} &= \frac{1}{2 \beta}\begin{pmatrix}
    -i \beta & 1 - i \alpha \\
    i \beta & 1 + i \alpha
  \end{pmatrix}.
\end{align}
Equation (\ref{eq:expAeta1}) can be rewritten as
\begin{equation*}
  \exp(\lambda_i \eta A) = 
  e^{\lambda_r \eta}
  \left(
    \cos(\lambda_i \eta)
    +
    \sin(\lambda_i \eta)
    V
    \begin{pmatrix}
      i & 0 \\
      0 & -i
    \end{pmatrix}
    V^{-1}
  \right).
\end{equation*}
Calculating
\begin{equation*}
V
\begin{pmatrix}
  i & 0 \\
  0 & -i
\end{pmatrix}
V^{-1}
\end{equation*}
and considering the starting point
$t'(0) = 0, x'(0) = x_s, z'(0) = z_s$,
we obtain
\begin{align}
  \begin{pmatrix}
    t'(\eta)\\z'(\eta)
  \end{pmatrix}
  &=
  z_s e^{\lambda_r \eta}
  \begin{pmatrix}
    0 & -\beta \\
    1 & \alpha
  \end{pmatrix}
  \begin{pmatrix}
    \cos {\lambda_i \eta} \\
    \sin {\lambda_i \eta}   
  \end{pmatrix} \label{eq:defspirtz}\\
  x'(\eta) &= x_s e^{-2 \lambda_r \eta} \label{eq:defspirx}
\end{align}
The equation (\ref{eq:defspirtz}) is an affine transformation of the logarithmic spiral:
\begin{equation}
  \begin{pmatrix}
    t''(\eta) \\ z''(\eta)
  \end{pmatrix}
  =  e^{\lambda_r \eta}
  \begin{pmatrix}
    \cos{\lambda_i \eta} \\
    \sin{\lambda_i \eta} 
  \end{pmatrix}
\end{equation}
This curve, named \emph{spira mirabilis} by Bernoulli three centuries ago,
is a spiral that never ends,
the first fractal curve ever studied.
For each loop toward the center, corresponding to a decrease in parameter $\eta$
equal to $-2 \pi/\lambda_i$,
the radius is decreased by a factor of $\exp(-2 \pi \lambda_r/\lambda_i) = 0.12$.
The spaghettis starting from a non-zero $x = x_s$ value
have the same $(t', z')$ behaviour,
spiralling toward $(t' = 0, z' = 0)$,
while they are escaping exponentially in $x$ direction,
with a factor
$\exp(4 \pi \lambda_r/\lambda_i) = \left(1/0.12\right)^2$
for each loop.

Despite this topological curiosity,
the electromagnetic field around the zero-field point
has a linear behaviour, as expected.
Indeed, the absolute \emph{characteristic parameter} $\eta$ we have introduced
has a constant value of $2 \pi/\lambda_i$ for each loop.
As the temporal and spatial
sizes of the loop are proportional to the distance to the center of the whirl,
the field amplitudes (Eq. (\ref{eq:fieldCompsFromEta})) are proportional
to this distance.

In order to control these computations,
the field components can be computed back from
absolute ES lines with $\eta$ parameter,
using Eqs.~(\ref{eq:fieldCompsFromEta}),
and  Eqs.~(\ref{eq:defspirtz}, \ref{eq:defspirx}), we calculate
\begin{align}
  \begin{pmatrix}
    H_y\\D_x
  \end{pmatrix}
  &=
  \frac{d}{d \eta}
  \begin{pmatrix}
    t'\\z'
  \end{pmatrix}
  \\
  D_z &= -\frac{d x'}{d \eta}
\end{align}
The component $D_z$ is the simplest to calculate :
We obtain $\lambda_x x_s e^{-\lambda_x \eta} = \lambda_x x'$
which is the expected value, given by Eq.~(\ref{eq:lindz}).
The two other components are
\begin{equation}
  \begin{pmatrix}
    H_y\\D_x
  \end{pmatrix}
  = z_s e^{\lambda_r \eta}
  \begin{pmatrix}
    -\lambda_i \beta & -\lambda_r \beta \\
    \lambda_r + \lambda_i \alpha & \lambda_r \alpha - \lambda_i
  \end{pmatrix}
  \begin{pmatrix}
    \cos{\lambda_i \eta} \\
    \sin{\lambda_i \eta} 
  \end{pmatrix}
\end{equation}
or
\begin{equation}
  \begin{pmatrix}
    H_y\\D_x
  \end{pmatrix}
  = z_s e^{\lambda_r \eta} \lambda_i \beta
  \begin{pmatrix}
    -1 & \alpha - \beta \\
    1 & \alpha - \beta
  \end{pmatrix}
  \begin{pmatrix}
    \cos{\lambda_i \eta} \\
    \sin{\lambda_i \eta}. 
  \end{pmatrix}
\end{equation}
Inverting Eq.~(\ref{eq:defspirtz})
we obtain
\begin{equation}
  z_s e^{\lambda_r \eta}
  \begin{pmatrix}
    \cos {\lambda_i \eta} \\
    \sin {\lambda_i \eta}   
  \end{pmatrix}
  =
  \frac{1}{\beta}
  \begin{pmatrix}
    \alpha & \beta \\
    -1 & 0
  \end{pmatrix}
  \begin{pmatrix}
    t'\\z'
  \end{pmatrix},
\end{equation}
and eventually
\begin{equation}
  \begin{pmatrix}
    H_y\\D_x
  \end{pmatrix}
  =  \lambda_i
  \begin{pmatrix}
    -2\alpha + \beta & - \beta \\
    \beta & \beta
  \end{pmatrix}
  \begin{pmatrix}
    t' \\
    z'
  \end{pmatrix},
\end{equation}
which is indeed the expected field component, given by Eq.~(\ref{eq:linhydx}).

\section{\label{sec:zefssol}Direct computation of ZEFSs}

As it is pointed out in the letter,
in a region of no electric charge the electric flux value between two ZEFSs is the same along any spaghetti segment.
If a $(O, \mathbf{t},\mathbf{x},\mathbf{z})$
frame of reference is chosen,
using the zero electric flux value property of a ZEFS,
the components of the electromagnetic fields between two ZEFS can be found as the ratio of $d\varphi_e$
over the distances $dt$, $dx$ and $dz$
in the directions $\mathbf{t}, \mathbf{x}, \mathbf{z}$
with the appropriate signs
\begin{equation}
     H_y = \frac{\partial \phi_e }{ \partial t},
     \;  D_z = \frac{\partial \phi_e }{\partial x},
     \;  D_x = -\frac{\partial \phi_e }{\partial z}.
    \label{eq:fromFluxFieldComp}
\end{equation}
Considering the $\phi_e$ potential
and Eq.~(\ref{eq:fromFluxFieldComp}) with the curl equation
$\partial_t B_y - \partial_x E_z + \partial_z E_x = 0$
considerably simplifies a $y$-invariant $p$-polarized electromagnetic problems with no charge.
In a vacuum, the resulting equation is just the d'Alembertian equal to zero:
$\left(\partial_t^2 - \partial_x^2 -\partial_z^2\right) \phi_e = 0$ with the normal derivative
Using this equation, ZEFSs can be built directly.

\section{Spaghetti construction}

ES and ZEFS can be build directly by solving the Maxwell's equations for any oriented 2D surface
$\Sigma$ of the 4D spacetime
\begin{align}
\Phi_m(\Sigma_\mathrm{closed}) &= 0\\
\Phi_e(\Sigma_\mathrm{closed}) &= n_e ,
\end{align}
where $n_e$ is the net number of electric charges
in any 3D volume enclosed by the 2D surface $\Sigma_\mathrm{closed}$, 
applying electric ($\bm{\tilde{\Phi}_e}$) and magnetic ($\bm{\tilde{\Phi}_m}$) two-forms to an infinitesimal oriented parallelogram
$dS$
\begin{align}
\Phi_m(\Sigma) &= \int_\Sigma \bm{\tilde{\Phi}_m}(dS)\\
\Phi_e(\Sigma) &= \int_\Sigma \bm{\tilde{\Phi}_e}(dS) .
\end{align}

The other way is based on the electromagnetic field components in a given reference.
In this section, the ES construction of Fig.~3(c) and Fig.~4 of the letter based on components of the electromagnetic field is described.
For Fig.~3 (d) a similar process is used except only ES were constructed for chosen starting points.

The description below is limited by the middle slit plane at $x=x_0$, as ES construction in any other plane is similar except that ESs can be directly built from the starting points without ZEFS.
As soon as the electromagnetic field components ($H_y(t,x,z)$, $D_x(t,x,z)$, $D_z(t,x,z)$) are found for the slit problem, the starting points for ZEFS $(s_{0t}, x_0, s_{0z})$ are defined in the slit region along the $t$ direction in the way that there is a unit electric flux step $\Delta \varphi$ between two neighbor points.
As in absence of an electric charge, a value of the electric flux is constant between two ZEFS, an increase / a decrease in the distance between them provides additional information on electric flux change.
ZEFS construction is based on its property of orthogonality to ES at a given event.
For the middle slit plane ($x= x_0$) the tangent vector to a ZEFS is
\begin{equation}
    \bm{\tau}_{Z} =  D_x \partial_t +  H_y \partial_z,
    \label{eq:tz}
\end{equation}
which defines the system of differential equations
\begin{eqnarray}
t'(\eta) = D_x (t(\eta),z(\eta)),  \\
z'(\eta) =  H_y (t(\eta),z(\eta)).
\label{eq:dez}
\end{eqnarray}

Then, ES is built as solution of the system of differential equations from Eq.~(6) of the letter.
As initial conditions for ESs, their property of coincidence with ZEFSs in the events of light lines is used.
In this case the system is supplemented by Eq.~(5) of the letter to provide the electric flux scale along the ES lines.

Several numerical techniques were used to compute
ESs and ZEFS for this paper:
computations of Fig.~3 and Fig.~4 were performed using the computer algebra software Wolfram Mathematica,
the file for Fig.~4 is in supplemented materials.
Both Wolfram Mathematica and Julia programming language were used to build Fig.~5.
The corresponding Mathematica and Julia-Pluto notebooks can be found respectively in repositories
\href{https://github.com/marinyakovleva/spaghetti}{https://github.com/marinyakovleva/spaghetti}
and \href{https://github.com/fp4code/arxiv-2208.11461}{https://github.com/fp4code/arxiv-2208.11461}.